\documentstyle[aps,multicol,epsf]{revtex} 
\begin{document} 

\title{
 Role of Defects in Self-Organized Criticality: A
Directed Coupled Map Lattice Model}

\author{
 Bosiljka Tadi\'c~$^{a,}$\cite{email1}  and
Ramakrishna Ramaswamy~$^{b,}$\cite{email2}} 

\address{
 $^a$ Jo\v{z}ef Stefan Institute, P.O. Box 100, 61111-Ljubljana,
Slovenia \\
$^b$ School of Physical Sciences, Jawaharlal Nehru University, 
New~Delhi 110 067, India}

\maketitle

\begin{abstract}

We study a directed coupled map lattice (CML) model in $d=2$ 
dimensions, with two
degrees of freedom associated with each lattice site.  The two freedoms
are coupled at  a fraction $c$ of lattice bonds acting as
quenched random defects.  The system is driven (by adding ``energy'', say)
in one of the degrees of freedom at the top of the lattice, and the
relaxation rules depend on the local difference between the two variables
at a lattice site. In the case of conservative dynamics, at any
concentration of defects the system reaches a self-organized critical
state with universal critical exponents close to the mean-field
values $\tau_t~=~1$, $\tau_s~=~2/3$, and $\tau_n~=~1/2$, for the 
integrated distributions of avalanche durations ($t$), size ($s$), 
and  released energy ($n$), respectively.  
The probability distributions follow the general scaling
form $P(X,L) = L^{-\alpha } {\cal{P}}(XL^{-D_X})$, where $\alpha \approx
1$ is the scaling exponent for the distribution of avalanche lengths, $X$
stands for $t$, $s$~ or $n$, and $D_X$ is the (independently determined)
fractal dimension with respect to $X$. The distribution of current through
the system is, however, nonuniversal, and does not show any apparent
scaling form.  In the case of nonconservative dynamics---obtained by
incomplete energy transfer at the defect bonds---the system is driven out
of the critical state. In the scaling region close to $c=0$ the 
probability distributions exhibit the   general
scaling form $P(X,c,L) = X^{-\tau _X }{\cal{P}}(X/\xi _X (c), XL^{-D_X})$,
where $\tau _X = \alpha /D_X$ and the corresponding coherence length $\xi
_X (c)$ depends on the concentration of defect bonds $c$ as $\xi _X (c)\sim
c^{-D_X}$.
\end{abstract}


\begin{multicols}{2}
\section*{I. Introduction}

Critical behavior in the vicinity of a second-order phase transition is
known to be sensitive to the presence of quenched random defects
\cite{Stinchcombe}. Quenched disorder can cause a change of the 
universality class of the critical behavior--in the case of weak disorder,
or lead to a variety of new phenomena, as in  random-field
systems \cite{rfs}.  The case of strong disorder, represented for example
by the presence of random bonds in a spin system, can lead to a
multiplicity of metastable states and frustration, as in spin glasses
\cite{sg}.

In contrast to conventional critical phenomena in systems which are tuned
to a critical point by varying one or more external parameters, much
recent work \cite{BTW} has examined the behavior of extended open
dynamical systems (sandpile cellular automata being a good example) which
tend to self-organize into a metastable states with long-range spatial and
temporal correlations. Such {\it self-organized criticality} (SOC) in a
system which is not tuned to a critical point can be expected to be
somewhat more robust and less sensitive to perturbations.

The question of relevance of such perturbations in SOC is of considerable
interest. A dynamic renormalization group study of spatially continuous
models \cite{Toner}, which are believed to describe some aspects of
self-organization found in cellular automata, shows that loss of
translational invariance due to quenched disorder is a relevant
perturbation for SOC. Similar considerations apply to the role of 
conservation laws in the evolution rules of such systems.  Various
examples studied so far \cite{noncons} reveal, however, that having a
conservation law in the dynamics is neither sufficient nor necessary
condition for the critical state to appear.

We explore some aspects of these questions in the present work, where we
introduce and study a {\it two-degree of freedom} (or {\it two-color})
directed coupled map lattice model on a 2-dimensional square lattice. In
our system there are two variables associated with each lattice
site, and these two freedoms are practically independent in the evolution,
except at a random fraction $c$~ of the lattice bonds, which act as
quenched random defects.  This model (termed Model B in earlier work
\cite{TR}) provides a simple example of a situation wherein quenched
disorder acts differently from annealed disorder \cite{T}.  Similar models
have also been studied in the context of signal transmission in a neural
network \cite{nn}. 

The system is driven---by adding ``energy'', say---to one of the degrees
of freedom at a random site at the top of the lattice. Instabilities can
be caused when the difference between the two variables at a site exceeds
a threshold value, in which case a relaxation occurs, and the total energy
accumulated in both states at an unstable site is transferred to the
forward neighboring sites, creating an avalanche.  Partitioning of the
energy between the states at neighboring sites depends on the kind of bond
(defect or normal) connecting these sites with the unstable neighbor (see
Section~II for details of the dynamics).  Avalanches can be characterized
through the usual indicators such as duration $t$, defined as number of
steps that the instability progresses toward the lower boundary of the
lattice, and size $s$---the number of lattice sites affected.  In
addition, we also monitor the total energy released in an avalanche (or
the number of relaxations) $n$, as well as the total number of particles
leaving the system at the lower boundary, {\it i.e.}~ the outflow current $J$.

Both conservative and nonconservative situations are possible by altering
the energy transfer at defect bonds. The present numerical simulations,
combined with a scaling analysis of the results help in determining the
conditions under which the coupled map lattice (CML) reaches a SOC state, 
and also the universality class to which the critical state belongs.  
When the dynamics is conservative, {\it i.e.}, the total energy 
which is removed from one site appears at its neighbors, 
the system self-organizes into a critical
state with  universal scaling exponents.  On the contrary, if the
energy transfer is incomplete at defect bonds, our results suggest that
the system is subcritical, with a finite coherence length which depends on
the defect concentration $c$.  We determine a new scaling function for the
distribution of size and duration of avalanches.

The case of site defects in a critical-height sandpile automaton has 
been examined in  previous work \cite{TR,TNURP}, where we showed that
the presence of nonconserving defects can lead to a loss of SOC. Site
defects (either annealed or quenched) introduce a coherence length in the
problem, which diverges as the concentration of defects vanishes; the
relevant exponents can be derived exactly \cite{Theiler}, since the
directed abelian sandpile with defects can be viewed as a random branching
process.

In Section II we introduce the model and define the scaling forms of
various distributions. Results are given in Section~III for conservative
dynamics and in Section~IV for the case of nonconservative dynamics,
followed by a short summary and discussion of the results, in Section~V.

\section*{II. Dynamical Model and Scaling}

The coupled-map lattice studied here is patterned on the
2-dimensional directed abelian sandpile cellular automaton \cite{DR} which
is a simple example of an exactly solvable system exhibiting SOC.  We
associate two dynamical variables $(h_1,h_2)$, which for convenience can
be termed energies, with each lattice site of a 2-dimensional directed
square lattice of linear size $L$. The relaxation process at lattice site
$(i,j)$ is determined by the actual values of $(h_1,h_2)$, as follows.  If
the absolute value of the {\it difference} between $h_1$ and $h_2$ exceeds
a critical value $d_c$,
{\it i.e.},
\begin{equation}
|h_1(i,j)-h_2(i,j)| \ge d_c \; , \label{instab}
\end{equation}\noindent
then the site $(i,j)$ becomes unstable and both $h_1$ and $h_2$ are reset
to zero,
\begin{equation}
h_1(i,j) \to 0 \; ; h_2(i,j) \to 0. \label{dr1}
\end{equation}\noindent
The lattice sites are connected by bonds which can be either positive or
negative, and these affect the subsequent relaxation rules differently.
(We consider the situation when most bonds are positive, and a fraction
$c$~ of negative bonds are distributed at random and act 
as defects.) Along positive bonds,

\begin{equation}
h_1(i+1,j_\pm ) \to h_1(i+1,j_\pm ) + (h_1(i,j)+h_2(i,j))/2,
\label{dr2}
\end{equation}\noindent
and along negative bonds
\begin{equation}
h_2(i+1,j_\pm ) \to h_2(i+1,j_\pm ) + (\lambda h_1(i,j)+h_2(i,j))/2.
\label{dr3}
\end{equation}\noindent 
$j_\pm =j\pm (1-(-1)^i)/2$ and $(i+1,j_\pm)$ label the two downstream
neighbors of $(i,j)$. As can be seen, for $\lambda =1$ the total energy
$h_1 + h_2$ disappearing from site $(i,j)$ reappears in either
$h_1$ or $h_2$ of the forward neighbors regardless of the parity of the
connecting bonds---the dynamics is conservative. If $\lambda \mathopen< 1$,
some amount of energy is lost along the negative bonds and the dynamics is
nonconservative. This case is considered in Section~IV below.

The system is driven by adding an energy unit at random
sites at the input at the top ({\it i.e.}, along the row $i = 1$)
\begin{equation}
h_1(1,j) \to h_1(1,j) + 1 \; ,
\end{equation}\noindent
and allowed to proceed following  the
rules embodied in Eqs.\ (\ref{instab}-\ref{dr3}) until there are no
further instabilities.  This constitutes an avalanche.  Starting from an
initially random configuration, the system eventually reaches a SOC state
\cite{TR,T,nn} when there are
avalanches of all duration and size scales. These can be characterized
through the following four quantities.

(a) The length, $\ell$, is the total distance that an avalanche
propagates.  $P(\ell) \sim \ell ^{-(1+\alpha )}$ is the probability
distribution of avalanches of length $\ell$.

(b) The size, $s$, is the number of sites at which relaxation occurs in one
avalanche. $D(s) \sim s^{-\tau _s }$ is the distribution of avalanches of
size $s$~ or greater.

(c) In sandpile automata distinction can be made between the number of 
particles toppled and the number of sites involved in an avalanche 
\cite{manna}. Similarly, here we note that the total amount of energy 
released in an avalanche, $n$, is not simply proportional to the size, $s$, 
and this introduces the distribution $Q(n) \sim n^{-\tau_n}$ for avalanches
of released energy $ \ge n$.

(d) The duration of an avalanche, $t$, is described by the 
distribution $P(t) \sim t^{-(1+\tau _t)}$.

From finite-size scaling arguments \cite{Ketal} these distributions
should obey the following general scaling form
\begin{equation}
P(X,L) = L^{-\alpha }{\cal{P}}(XL^{-D_X}) ,
\label{scalgen}
\end{equation}\noindent
where $\alpha $ is the above defined exponent of the distribution of
lengths $P(\ell)$, and $X$ represents $s$, $n$~ or $t$.  Correspondingly,
$D_X$ stands for the appropriate fractal dimension, which is defined via
following relations:
\begin{equation}
\mathopen<s\mathclose>_\ell \sim \ell^{D_s} \; , \label{DFS}
\end{equation}\noindent
\begin{equation}
\mathopen<n\mathclose>_\ell \sim \ell^{D_n} \; , \label{DFN}
\end{equation}\noindent
and
\begin{equation}
\mathopen<t\mathclose>_\ell \sim \ell^{D_t} ,  \label{DFZ}
\end{equation}\noindent
where $\mathopen< \mathclose>_\ell $ denotes an average over all
avalanches of selected length $\ell$ ({\it i.e.}, avalanches which 
exactly terminate on the $\ell$'th row).

Note that $D_t$ is the dynamic exponent (usually denoted $z$). In the
present directed model, since avalanches propagate only in the forward
direction, the duration is equivalent to the length. The dynamic exponent
is thus exactly 1, and $\tau _t \equiv \alpha $. This need not be the case
when the relaxation rules are more complex \cite{TR}.  Furthermore, all
the exponents are not independent, since the scaling relations \cite{TR}
\begin{equation}
\alpha  = D_s\ \tau _s =D_n\ \tau _n =\tau _t\ D_t \; , 
\label{SCR}
\end{equation}\noindent
are valid for the exponents of integrated distributions as defined above.

In the present work we consider quenched disorder, {\it i.e.}, 
our numerical results are averaged over several sample lattices, 
each prepared by distributing defect bonds with concentration $c$. 
We keep the lattice
configuration fixed for a large number of Monte Carlo steps (which is
equal to number of events), and consider both cases of conservative and
nonconservative transfer at defect bonds.  In order to minimize effects of
boundaries, we chose so-called free boundaries in the perpendicular
direction. This is achieved by using a lattice of size $2L\times L$ and
initializing avalanches between the sites $L/2 +1$ and $3L/2$ at top.
According to the above dynamic rules, only escape of energy is possible at
bottom of the lattice between sites 1 and $2L$.  Results of our numerical
studies are presented in Sections III and IV.

\section*{III. Conservative Dynamics}

For $\lambda = 1$ in Eq.\ \ref{dr3}, the dynamics is conservative.  The
system reaches the SOC state, which we characterize through the quantities
enumerated in Section~II. It is sufficient to consider the disorder regime
of $c \le 0.5$ owing to the symmetry $c \to 1 - c, h_1 \to h_2$. When $c =
0$ this model reduces to a directed abelian sandpile automaton in the
$h_1$ degree of freedom, and the empty state in the other degree of
freedom, $h_2 = 0$ at all sites.  For nonzero $c$, the SOC state is more
difficult to describe.  We study the hystogram of total energy per site,
defined as $E=h_1+h_2$  after a relaxation event,
both for $c=0$ and few values of $c\neq 0$. In the case $c=0$ it takes
nonzero values at the interval $[0,1]$, however, for $c\neq 0$ much larger
values of energies occur, although with smaller probability compared to
those at $E=0$ and $E=1$. 
Spreading of the distribution in the presence of defect bonds $c\neq 0$
indicates that the critical state is realized via multiplicity of
configurations of the dynamic variables $h_1$ and $h_2$, which depends on
$c$.

We observe however, that the concentration of defect bonds $c\neq 0$
does {\it not} appear to affect the critical exponents characterizing 
the SOC state in this model.  
The distribution of lengths of the relaxation clusters $P(\ell)$, 
size $D(s)$ and number of relaxations $Q(n)$ for two different
concentrations of defect bonds $c = 0.1$ and $c = 0.5$ are shown in
Fig.~1.  The exponents are numerically determined to be $\alpha = 1.051
\pm 0.044$, $\tau _s = 0.650 \pm 0.028$, and
$\tau _n = 0.509 \pm 0.004$, {\it independent} of defect concentration
$c$, suggesting that this model has universal criticality. 
 
In Fig.\ 1d are shown the distributions for length and size of 
avalanches  $P_0(\ell )$ and $D_0(s)$ are shown for  the case $c=0$,. 
Slopes of these curves,  corresponding to the above defined exponents 
$1+ \alpha $ and $1+\tau _s$, 
respectively, are estimated as $1.998 \pm 0.019$ and $1.58\pm 0.04$. 
For $\tau _n $ we find the value $0.528 \pm 0.022$.

In Fig.\ 2 we  present
results for the time averaged  size $\mathopen<s\mathclose>$ and 
relaxed energies $\mathopen<n\mathclose>$ in avalanches
of a fixed (selected) length $\ell$. According to Eqs.\ 
(\ref{DFS})-(\ref{DFN}), slopes of these curves determine the 
mass-to-scale ratios (fractal dimensions) $D_s$ and $D_n$, which  
are determined to be $D_s = 1.615 \pm 0.007$ and $D_n
=1.997 \pm 0.008$. Plotted are the results for $c=0.5$, but it was 
checked that the values of the exponents remain concentration 
independent.

In order to fully characterize the self-organized critical state, we have
determined the various exponents and the corresponding scaling functions.
The results of the finite-size scaling fit according to Eq.\
(\ref{scalgen}) are shown in Figs.\ 3 and 4, where we used $\alpha
=1.04$, $D_s=1.62$ and $D_n=2$.

In contrast to many sandpile models, where the number of topplings 
(corresponding to $n$ in our model) is
expected to scale with a fractal exponent \cite{JC}, we find that the
relaxation rules (\ref{instab})-(\ref{dr3}) lead to rather classical
exponent $D_n =2$. The distributions of the size of relaxation clusters
$D(s,L)$ and the distribution of number of relaxations $Q(n,L)$ satisfy
the finite-size scaling form (\ref{scalgen}) with the above
determined exponents.

The distribution of outflow current $G(J,L)$, which is the current which
flows over the lower boundary of the system, rather than having a
power-law dependence on $J$, fulfills the finite-size scaling form
\cite{Ketal}
\begin{equation}
G(J,L) =L^{-\beta }{\cal{G}}(JL^{-\phi }) \; , \label{Current-sc}
\end{equation}\noindent
where $\beta = 2\phi $ in the stationary state. In Fig.\ 5 we show the
distribution $G(J,c,L)$ for (a) fixed concentration of defect bonds $c =
0.5$ and various lattice sizes, and (b) for fixed lattice size $L = 192$
and various concentrations of defects $c$.  The distribution $G(J,c,L)$ for
finite concentration $c$~ of defect bonds is rather localized, as opposed
to the case $c = 0$, where the distribution is broad.  We find no apparent
finite-size scaling for $G(J,c,L)$ for nonzero values of $c$ .

We end this section with a comment on numerical values of the critical
exponents.  Our results suggest that the values of the exponents for
avalanche duration and for number of relaxations are close to  those in
the mean-field SOC \cite{Zapperietal,MFSOC}--- in our notation, $1+\tau _t
\approx 2$, and $1+ \tau _n \approx 3/2$. The mean-field universality 
class \cite{Zapperietal} is obtained in self-organizing sandpile
models where: (i) there is a front-like spreading of avalanches with
noninteracting sites at the front, and (ii) a global constraint exists, 
which maps the dynamic model to a {\it critical } branching
process. Although our model has somewhat more complex evolution rules,
the exponents appear to be in the same universality class.  (Note
that in our model, compared to the sandpile automata, there is one 
more independent exponent, {\it i.e.}, $1+\tau _s \approx 5/3$.)

\section*{IV. The Case of Nonconservative Defects}

When $\lambda \neq 1$ in Eq.\ (\ref{dr3}) the model becomes
nonconservative at defects bonds. By studying the probability 
distributions of  duration $P(t,c,L)$ and size $D(s,c,L)$ of 
avalanches for various concentrations of nonconserving defect bonds 
$c$, we find that SOC behavior is lost as soon as energy 
conservation is lost.

One quantity which proves to be useful in characterizing the present
situation is the average number of relaxations $\mathopen<n(\ell
)\mathclose>$ which occur in all kinds of clusters up to length $\ell $
(which is different from the average number of relaxations in clusters of
selected length, which was considered above). In the case of conservative
dynamics this quantity exhibits a power-law \cite{TR}
\begin{equation}
\mathopen<n(\ell )\mathclose> \sim \ell ^{D_n(1-\tau _n)} \; , \label{nl}
\end{equation}\noindent
independent of the degree of disorder.  

Numerical results are obtained by fixing $\lambda =0.9$ and varying $c$.
Shown in Fig.\ 6 are results for the average number of topplings
$\mathopen<n(\ell)\mathclose>$, for the conservative case (open circles)
as well as for $\lambda \mathopen< 1$ and different values of $c$, where
it can be seen that the power-law is lost and the results depend on the
concentration of defect bonds $c$~ (the remaining lines in Fig.\ 6).  

We further study the effects of concentration of defect bonds on the
distributions of size and duration (or length) of the relaxation clusters.
In Figs.\ 7 a and b these distributions are shown for $L = 128$ and for
few values of $c$. The following general scaling form is appropriate 
for the case of lattice disorder \cite{TNURP}

\begin{equation}
P(X,c,L) = X^{-\tau _X }{\cal{P}}(X/\xi _X(c), X/L^{D_X}) \; .
\label{ptc}
\end{equation}\noindent
Here $\xi _X(c)$ is the corresponding coherence length, which, as is
evident in Fig.\ 6, varies with the concentration of defect bonds $c$.  
By fitting the numerical data in Fig.\ 7 a and b to the scaling form
(\ref{ptc}) we determine the $c$-dependence of $\xi _X(c)$, which 
appears to be most satisfactorily described as $\xi _t= 1/c$ and 
$\xi _s =c^{-D_s}$, for the distributions of duration and size, 
respectively, $\tau _t\equiv \alpha $, $\tau _s $, and $D_s$ being 
the exponent determined in Sec.\ III.  The results are shown
in Fig.\ 7 c and d. Owing to the factor $t/L$ or $s/L^{D_s}$ in the
scaling function (\ref{ptc}), we restrict the fits to rather large
concentrations $c$ (see  Fig.\ 7 a and b) in order to satisfy the
relation $\xi /L \ll 1$.  It is interesting to note that the scaling 
region where the Eq.\ (\ref{ptc}) applies is restricted to a finite 
range of values of $c \mathopen< c^*$ in the vicinity of 
the point $c=0$. For instance, the curves
corresponding to $c=0.5$ in Fig.\ 7a, b do not obey the scaling form
(\ref{ptc}), indicating that $c^* \mathopen< 0.5$.
In the limit  $c=0$ the dynamics becomes 
conservative and true SOC reappears (cf. Fig.\ 1d.).

Similar scaling fits were introduced earlier in
Ref.~\cite{TNURP} in a simple critical height model with site defects.
Here we demonstrate that in the case of more complicated relaxation 
rules with nonconservative bond defects, the coherence lengths have 
the same general dependence on the concentration of defects, namely, 
$\xi _X(c) \sim c^{-D_X}$, where, in principle, the values of the 
exponents $D_X$  depend on the dynamic model.  Exact expressions 
for the coherence lengths in the case of the directed abelian sandpile 
model with site defects have been obtained by mapping the model to a 
random branching process \cite{Theiler}.

\section*{V. Summary and Discussion }

In this work we have further explored the role of defects in models of
self-organized criticality. In the case of sandpiles with random site
defects \cite {TNURP}, the dynamics is altered locally at defect sites.
The situation of defects having a more global influence is also of
interest, and we achieve this in the present two-color random bond model.
Our coupled map lattice has two degrees of freedom associated with each
lattice site. Disorder is present in the form of quenched random bond
defects, which can be both conserving ($\lambda = 1$) or nonconserving 
($\lambda \ne 1$). There is a preferred direction of transport, which 
serves to simplify the dynamical evolution rules, leading to a minimal 
model with quenched random bonds.  Preliminary studies \cite{TR,T} 
have indicated that annealed and quenched random disorder can behave 
differently, and our model is one of the simplest examples. Similar 
self-organizing coupled map lattice models have been studied to some 
extent in the literature as models of more realistic sandpiles \cite{FAO}, 
abstract examples of adaptive self-organizing systems \cite{Sinha}, 
vector-state models \cite{Peng}, and as nondirected neural networks 
\cite{nn}.

With conservation, $\lambda = 1$, in the limit of no disorder, $c = 0$,
the model becomes an abelian CML with single dynamic variable (energy
$h_1$), and reaches a SOC state in which the distribution $D(h_1)$ is
nonzero at the interval $[0,1]$, and $h_2 = 0$ everywhere. For nonzero
concentration of defect bonds ($c \neq 0$), additional state $h_2$ is
generated at fraction $c$ of lattice sites, which affects the propagation
of avalanches. The SOC state is again reached under the condition that the
total energy (consisting of $h_1+h_2$) accumulated at an unstable site is
completely transferred to its neighbors. Conservation of $h_1$ by itself
appears to be insufficient. The critical state has more complex structure,
evidenced by a spread in the distribution of the values of $h_1$ and
$h_2$, the width of the distribution being a strong function of $c$.
However, the critical {\it exponents} characterizing the SOC state appear
to be {\it independent} (within numerical uncertainty) of the
concentration of defects $c$.  Numerical values of these exponents suggest
that both $c = 0$ and $c \mathclose> 0$ models belong to the mean-field
SOC universality class \cite{Zapperietal,MFSOC}.  In particular, exponents
for the distributions of duration (or length) of relaxation clusters and
number of relaxations are close to the exact values $1+\alpha
\approx 2$ and $1+\tau _n \approx 3/2$, respectively.
Due to the more complex dynamical rules in our CML model when $c \neq 0$, 
we can distinguish between $\tau _n$ and the exponent of the distribution 
of size of avalanches, $\tau _s$, which are equivalent in abelian sandpile 
models \cite{manna}; the present results suggest that $1+ \tau _s \approx 
5/3$.

By modifying the relaxation rules in a way such as to permit incomplete
energy transfer at defect bonds, ({\it i.e.}, $h_2$ is not conserved 
while $h_1$ remains conserved), we study the significance of the 
conservation on the SOC state.  Similar to the case of sandpile automata 
with site defects, the self-organizing CML is driven out of the critical 
state (cf. Fig.\ 7), with the concentration of defect bonds $c$ playing 
the role of a control parameter. 
The coherence lengths are  tuned by the concentration of defect bonds 
according to the general law $\xi _X(c) \sim c^{-D_X}$. In
the case of site defects, the sandpile automaton can be mapped onto a
random branching process, and the exact expression for the $\xi _t(c)$ 
has been derived (see Ref. \cite{Theiler} for details).
We suggest that the scaling form (\ref{ptc}) together with the 
expression $\xi _X(c) \sim c^{-D_X}$ for the coherence lengths applies 
for the probability distributions in a wider class of {\it subcritical} 
self-organizing systems.

In contrast to spatially continuous models in which disorder is always a
relevant perturbation \cite{Toner}, our numerical results suggest that the
present coupled-map lattice model self-organizes into a universality class
of the mean-field SOC which is robust to quenched random bonds, as long as
the dynamics is conservative.  However, a question remains as if 
other types of defects, such as studied in discrete sandpile automata in
Refs.\ \cite{TR,LTU}, combined with different dynamic rules may
lead to continuous tuning of the universality class with a parameter in
CML models.

\section*{Acknowledgments} 
This work was supported by the Ministry of Science and Technology of the
Republic of Slovenia, and the DST, India under grant SP/S2/MO5/92.

\end{multicols}

\begin{figure}
\caption{ Double 
logarithmic plot of (a) distribution of avalanche lengths $P(\ell )$
vs. $\ell $; (b) integrated distribution of avalanche sizes $D(s)$ vs. 
$s$; (c) integrated distribution of number of relaxations $Q(n)$ vs. 
$n$, for two concentration of defect bonds $c=0.1$ (dotted curves)
and $c=0.5$ (solid curves); (d) distributions of avalanche lengths 
and sizes (not integrated), $P_0(\ell )$ vs. $\ell $ and $D_0(s)$ vs. 
$s$, respectively, for the case $c=0$.
\label{fig:1}}
\end{figure}

\begin{figure}
\caption{Double-logarithmic plot 
of the average values of size $\mathopen<s\mathclose>_\ell $ and
number of relaxations $\mathopen<n\mathclose>_\ell $  in clusters of
selected length $\ell $, vs. $\ell $.
\label{fig:2}}
\end{figure}

\begin{figure}
\caption{Double-logarithmic plot of the integrated distribution of 
size of avalanches $D(s,L)$ vs. size $s$ for several different values 
of lattice size $L$, (above), and the corresponding finite-size 
scaling plot according to Eq.\ (6),  (below).
\label{fig:3}} 
\end{figure}

\begin{figure}
\caption{Same as Fig.\ 3 but for the integrated distribution of 
number of relaxations $Q(n,L)$ vs.  $n$ for three different values of
lattice size $L$, (top), and the corresponding finite-size-scaling plot
according to Eq.\ (6), (bottom).
\label{fig:4}}
\end{figure}

\begin{figure}
\caption{Double-logarithmic plot 
of the distribution of outflow current $G(J,c,L)$ vs. current $J$
for fixed concentration of defect bonds $c=0.5$ 
and various lattice sizes $L$, (top), and for fixed $L=192$ and various 
concentrations $c$ as indicated, (bottom).
\label{fig:5}}
\end{figure}

\begin{figure}
\caption{Double-logarithmic plot 
of the average number of relaxations up to distance $\ell $ from 
the top of pile  $\mathopen<n(\ell )\mathclose>$ vs. $\ell $ for 
conservative defects $\lambda =1$ (open circles) and for $\lambda 
\mathopen< 1$ and various concentrations of nonconservative defects 
$c$ as indicated.
\label{fig:6}}
\end{figure}

\begin{figure}
\caption{Top: Double-logarithmic plot of the distribution of length 
$P(\ell , c)$ vs. length $\ell $ and the integrated distribution of 
size $D(s,c)$ vs. size $s$ of avalanches  for few concentrations of 
nonconservative defects (from right to left: $c=$ 0.1, 0.2, 0.3, and 
0.5) and fixed $L=$ 128. Bottom: Scaling plots of the same 
distributions for $c=$0.1, 0.2, and 0.3 according to Eq.\ (13).
\label{fig:7}}
\end{figure}

\end{document}